\documentclass[a4paper]{article}

\usepackage{ISCSLP_v2}
\usepackage{multirow}
\usepackage{graphicx}
\usepackage{subcaption}

\title{Angular Softmax Loss for End-to-end Speaker Verification}
\name{Yutian Li, Feng Gao, Zhijian Ou, Jiasong Sun\thanks{This work is supported by NSFC grant 61473168. Correspondence to: Z. Ou (ozj@tsinghua.edu.cn).}}
\address{
  Speech Processing and Machine Intelligence (SPMI) Lab, Tsinghua University, Beijing, China}
\email{yutian-l16@mails.tsinghua.edu.cn, ozj@tsinghua.edu.cn}

\begin{document}

\maketitle
\begin{abstract}	
End-to-end speaker verification systems have received increasing interests.
The traditional i-vector approach trains a generative model (basically a factor-analysis model) to extract i-vectors as speaker embeddings.
In contrast, the end-to-end approach directly trains a discriminative model (often a neural network) to learn discriminative speaker embeddings; a crucial component is the training criterion.
In this paper, we use angular softmax (A-softmax), which is originally proposed for face verification, as the loss function for feature learning in end-to-end speaker verification. 
By introducing margins between classes into softmax loss, A-softmax can learn more discriminative features than softmax loss and triplet loss, and at the same time, is easy and stable for usage.
We make two contributions in this work. 1) We introduce A-softmax loss into end-to-end speaker verification and achieve significant EER reductions. 2) We find that the combination of using A-softmax in training the front-end and using PLDA in the back-end scoring further boosts the performance of end-to-end systems under short utterance condition (short in both enrollment and test).
Experiments are conducted on part of $Fisher$ dataset and demonstrate the improvements of using A-softmax.
	
\end{abstract}

\section{Introduction}

Speaker verification is a classic task in speaker recognition,
which is to determine whether two speech segments are from the same speaker or not. 
For many years, most speaker verification systems are based on the i-vector approach \cite{dehak2011front}.
The i-vector approach trains a generative model (basically a factor-analysis model) to extract i-vectors as speaker embeddings, and relies on variants of probabilistic linear discriminant analysis (PLDA) \cite{prince2007probabilistic} for scoring in the back-end.

End-to-end speaker verification systems have received increasing interests.
The end-to-end approach directly trains a discriminative model (often a neural network) to learn discriminative speaker embeddings.
Various neural network structures have been explored. 
Some studies use RNNs to extract the identity feature for an utterance \cite{zhang2016end}\cite{heigold2016end}\cite{chowdhury2017attention}\cite{li2017deep}\cite{bredin2017tristounet}. 
Usually, the output at the last frame from the RNN is treated as the utterance-level speaker embedding. Various attention mechanisms are also introduced to improve the performance of RNN-based speaker verification systems. 
There are also some studies based on CNNs \cite{zhang2016end}\cite{li2017deep}\cite{lilantian2017deep}\cite{torfi2017text}\cite{zhang2017end},
where the f-bank features are fed into the CNNs to model the patterns in the spectrograms.

\begin{figure}[t]
	\centering
	\begin{subfigure}[b]{0.5\textwidth}
		\centering
		\includegraphics[width=\linewidth]{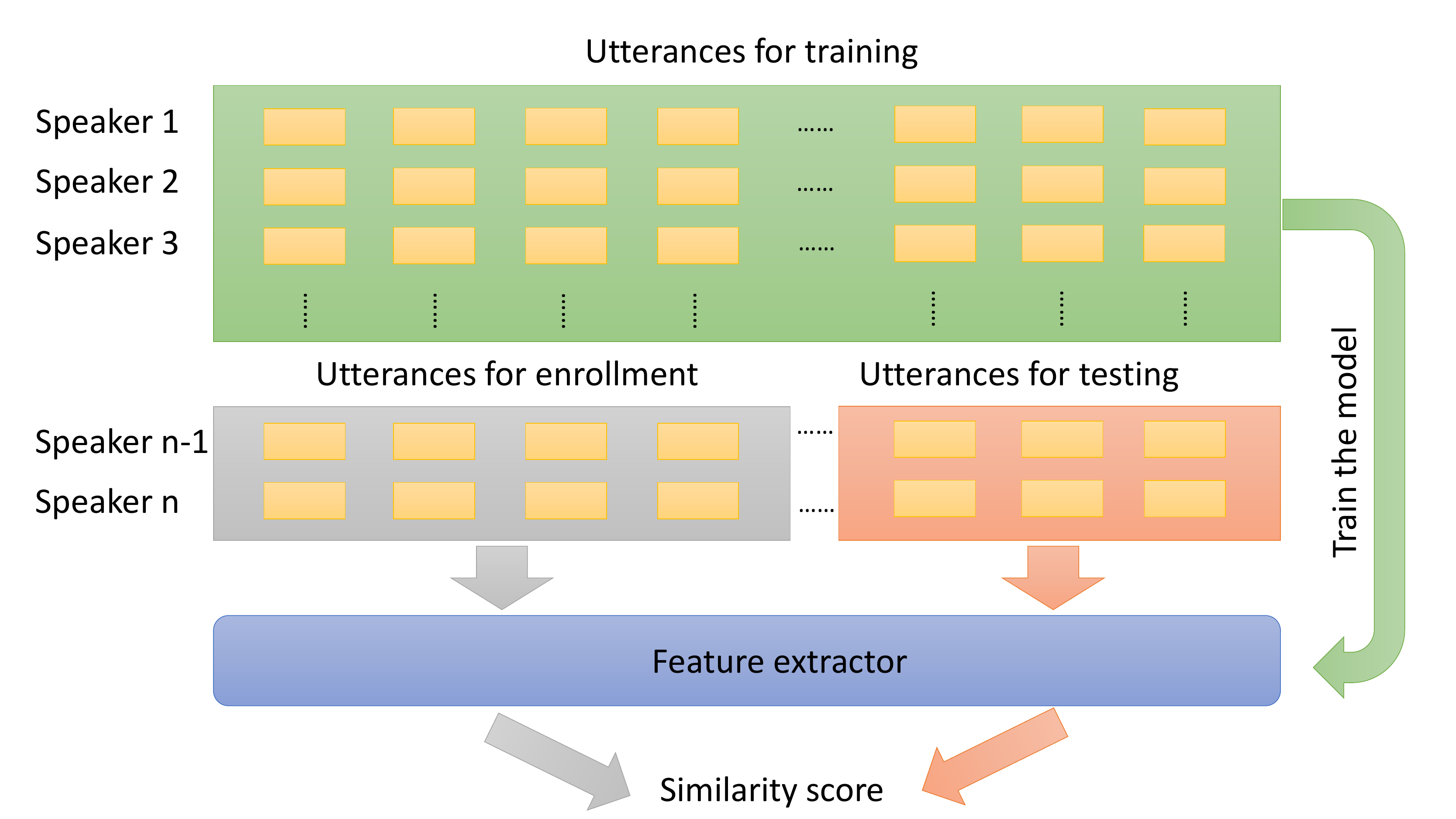}
		\caption{Verification task}
	\end{subfigure}
	\begin{subfigure}[b]{0.5\textwidth}
		\centering
		\includegraphics[width=\linewidth]{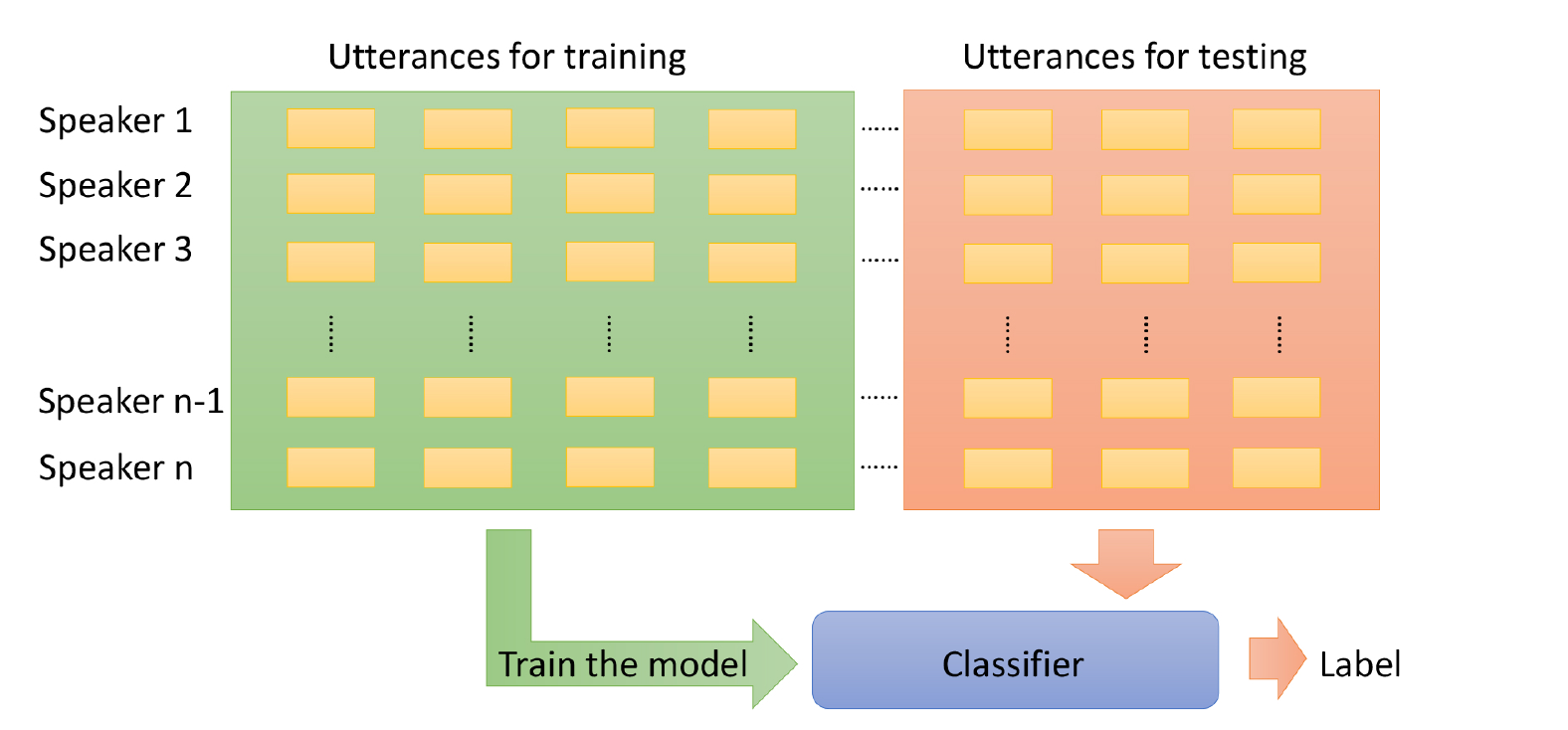}
		\caption{Classification task}
	\end{subfigure}
	\caption{
	    Speech Utterances are represented by small yellow blocks.
	    Each row represents a speaker with a number of utterances.
	    The above illustrates the verification task, which consists of feature extraction and scoring.
		The speakers used for training the feature extractor usually do not appear in testing.
	    The bottom shows the classification task. The speakers used for training the classifier appear in testing. Namely, in testing, utterances from the same set of training speakers are presented for classification. 
    }
	\label{fig:1}
	\vskip -0.2in
\end{figure}

In addition to exploring different neural network architectures, an important problem in the end-to-end approach is to explore different criteria (loss functions), which drive the network to learn discriminative features. 
In early studies, the features extracted by the neural networks are fed into a softmax layer and the cross entropy is used as the loss function. This loss is generally referred to as ``softmax loss''.
But the softmax loss is more suitable for classification tasks (classifying samples into given classes). 
In contrast to classification, verification is an open-set task. Classes observed in the training set will generally not appear in the test set. Figure \ref{fig:1} shows the difference between the classification and verification tasks. 
A good loss for verification should push samples in the same class to be closer, and meanwhile drive samples from different classes further away. 
In other words, we should make inter-class variances larger and intra-class variances smaller. 
A number of different loss functions have been proposed to address this problem \cite{li2017deep}\cite{wan2017generalized}\cite{snyder2016deep}.

Triplet loss \cite{schroff2015facenet} is recently proposed to take inter-class and intra-class variances into consideration. 
Triplet loss based training requires a careful triplet selection procedure, which is both time-consuming and performance-sensitive.
There are some interesting efforts to improve the triplet loss based training, such as generating triplets online from within a mini-batch \cite{schroff2015facenet}, doing softmax pre-training \cite{li2017deep}.
However, training with the triplet loss remains to be a difficult task.
Our experiment of using triplet loss yields inferior performance, compared to the i-vector method.

Angular softmax (A-softmax) loss \cite{liu2017sphereface} is recently proposed to improve the softmax loss in face verification. It enables end-to-end training of neural networks to learn angularly discriminative features.
A-softmax loss introduces a margin between the target class and the non-target class into the softmax loss. 
The margin is controlled by a hyper-parameter $m$. The larger $m$ is, the better the network will perform. Compared with the triplet loss, A-softmax is much easier to tune and monitor.

In this paper, we introduce A-softmax loss into end-to-end speaker verification, as the loss function for learning speaker embeddings.
In \cite{liu2017sphereface}, cosine distance is used in the back-end scoring.
Beyond of this, we study the combination of using A-softmax in training the front-end and using PLDA in the back-end scoring.
Experiments are conducted on part of the $Fisher$ dataset.
The neural network structure is similar to that used by the Kaldi xvector \cite{snyder2017deep}.
Using A-softmax performs significantly better than using softmax and triplet loss.
The EERs of A-softmax system are the best on almost all conditions, except that both the enroll and the test utterances are long. 
It is known that the i-vector based system performs well under such long utterance condition \cite{snyder2016deep,lilantian2017deep}. 
We also find that under short utterance condition (short in both enrollment and test), using PLDA in the back-end can further reduce EERs of the A-softmax systems.

\section{Method}

A-softmax loss can be regarded as an enhanced version of softmax loss. The posterior probability given by softmax loss is:
\begin{displaymath}
	p_i = \frac{e^{\mathbf{W}_i^T\mathbf{x} +b_i}}{\sum_{j}e^{\mathbf{W}_j^T\mathbf{x}+b_j}}
\end{displaymath}
where $\mathbf{x}$ is the input feature vector. 
$\mathbf{W}_i$ and $b_i$ are the weight vector and bias in the softmax layer corresponding to class $i$, respectively.

To illustrate A-softmax loss, we consider the two-class case.
It is trivial to generalize the following analysis to multi-class cases.
The posterior probabilities in the two-class case given by softmax loss are:
\begin{displaymath}
p_1 = \frac{e^{\mathbf{W}_1^T\mathbf{x}+b_1}}{e^{\mathbf{W}_1^T\mathbf{x}+b_1}+e^{\mathbf{W}_2^T\mathbf{x}+b_2}}
\end{displaymath}
\begin{displaymath}
p_2 = \frac{e^{\mathbf{W}_2^T\mathbf{x}+b_1}}{e^{\mathbf{W}_1^T\mathbf{x}+b_1}+e^{\mathbf{W}_2^T\mathbf{x}+b_2}}
\end{displaymath}
The predicted label will be assigned to class 1 if $p_1 \ge p_2$ and class 2 if $p_1 < p_2$.
The decision boundary is $\,(\mathbf{W}_1^T-\mathbf{W}_2^T)\,\mathbf{x}=0$, which can be rewritten as $(\,\lVert\mathbf{W}_1\rVert\cos(\,\theta_1)\,-\lVert\mathbf{W}_2\rVert\cos(\,\theta_2)\,)\,\lVert \mathbf{x} \rVert=0$. 
Here $\theta_1$, $\theta_2$ are the angles between $\mathbf{x}$ and $\mathbf{W}_1$, $\mathbf{W}_2$ respectively.

There are two steps of modifications in defining A-softmax \cite{liu2017sphereface}.
First, when using cosine distance metric, it would be better to normalize the weights and and zero
the biases, i.e. $\lVert\mathbf{W}_1\rVert = \lVert\mathbf{W}_2\rVert = 1$ and $b_1=b_2=0$. 
The decision boundary then becomes angular boundary, as defined by $\cos(\,\theta_1)\,-\cos(\,\theta_2)\,=0$.
However, the learned features are still not necessarily
discriminative.
Second, \cite{liu2017sphereface} further proposes to incorporate angular margin to enhance the discrimination power. Specifically, an integer $m$ ($m \ge 2$) is introduced to quantitatively control the size of angular margin.
The decision conditions for class 1 and class 2 become $\cos(\,m\theta_1)\,-\cos(\,\theta_2)\,>0$ and $\cos(\,m\theta_2)\,-\cos(\,\theta_1)\,>0$ respectively.
This means when $\cos(\,m\theta_1)>\cos(\,\theta_2)\,$, we assign the sample to class 1; when  $\cos(\,m\theta_2)>\cos(\,\theta_1)\,$, we assign the sample to class 2.

Such decision conditions in A-softmax are more stringent than in the standard softmax.
For example, to correctly classify a sample $\mathbf{x}$ from class 1, A-softmax requires $\cos(m \theta_1) > \cos(\theta_2)$, which is stricter than $\cos(\theta_1) > \cos(\theta_2)$ as required in the standard softmax.
Because that the cosine function is monotonically decreasing in $[0,\pi]$, when $\theta_1$ is in $[0,\frac{\pi}{m}]$, we have $\cos(\theta_1) > \cos(m\theta_1) > \cos(\theta_2)$. 
It is shown in \cite{liu2017sphereface} that when all training samples are correctly classified according to A-softmax, the A-softmax decision conditions will produce an angular margin of $\frac{m-1}{m+1}\varTheta$, where $\varTheta$ denotes the angle between $\mathbf{W}_1$ and $\mathbf{W}_2$.

By formulating the above idea into the loss function, we obtain the A-softmax loss function for multi-class cases:
\begin{displaymath}
	L = \frac{1}{N}\sum_{n=1}^{N}-\log\frac{e^{\lVert\mathbf{x}^{(n)}\rVert\cos(\,m\theta^{(n)}_{y_n})\,}}{e^{\lVert\mathbf{x}^{(n)}\rVert\cos(\,m\theta^{(n)}_{y_n})\,}+\sum_{j\ne y_n}e^{\lVert\mathbf{x}^{(n)}\rVert\cos(\,\theta^{(n)}_j)\,}}
\end{displaymath}
where $N$ is the total number of training samples. 
$\mathbf{x}^{(n)}$ and $y^{(n)}$ denote the input feature vector and the class label for the $n$-th training sample respectively.
$\theta^{(n)}_j$ is the angle between $\mathbf{x}^{(n)}$ and $\mathbf{W}_j$, and thus $\theta^{(n)}_{y_n}$ denotes the angle between $\mathbf{x}^{(n)}$ and the weight vector $\mathbf{W}_{y_n}$.

Note that in the above illustration, $\theta^{(n)}_{y_n}$ is supposed to be in $[0,\frac{\pi}{m}]$. 
To remove such restriction, a new function is defined to replace the cosine function as follows:
\begin{displaymath}
	\phi(\,\theta^{(n)}_{y_n})\, = (\,-1)\,^k\cos(\,m\theta^{(n)}_{y_n})\,-2k
\end{displaymath}
for $\theta^{(n)}_{y_n} \in[\frac{k\pi}{m},\frac{(k+1)\pi}{m}]$ and $k\in[0,m-1]$. 
So the A-softmax loss function is finally defined as follow:
\begin{displaymath}
	L = \frac{1}{N}\sum_{n=1}^{N}-\log\frac{e^{\lVert\mathbf{x}^{(n)}\rVert\phi(\,\theta^{(n)}_{y_n})\,}}{e^{\lVert\mathbf{x}^{(n)}\rVert\phi(\,\theta^{(n)}_{y_n})\,}+\sum_{j\ne y_n}e^{\lVert\mathbf{x}^{(n)}\rVert\cos(\,\theta^{(n)}_j)\,}}
\end{displaymath}

By introducing $m$, A-Softmax loss adopts different
decision boundaries for different classes (each boundary
is more stringent than the original), thus producing angular
margin.  The angular margin increases with larger $m$ and would be zero if $m=1$.
Compared with the standard softmax, A-softmax makes
the decision boundary more stringent and separated, which can drive more discriminative feature learning.
Compared with the triplet loss, using A-softmax loss do not need to sample triplets carefully to train the network.

Using A-softmax loss in training is also straightforward. 
During forward-propagation, we use normalized network weights.
To facilitate gradient computation and back-propagation,
$\cos(\theta^{(n)}_j)$ and $\cos(m\theta^{(n)}_{y_n})$ can be replaced by expressions only containing $\mathbf{W}$ and $\mathbf{x}^{(n)}$, according to the definition
of cosine and multi-angle formula\footnote{That is the reason why	we need $m$ to be an integer.}. In this manner, we can compute
derivatives with respect to $\mathbf{W}$ and $\mathbf{x}^{(n)}$, which is similar to using softmax loss in training.


%

\section{Experimental Setup}

This section provides a description of our experimental setup including the data, acoustic features, baseline systems and the neural network architectures used in our end-to-end experiments.
We evaluate the traditional i-vector baseline and Kaldi xvector baseline \cite{snyder2016deep,snyder2017deep}, which is an end-to-end speaker verification system recently released as a part of Kaldi toolkit \cite{Povey_ASRU2011}.
We also conduct triplet loss experiments for comparison.

\subsection{Data and acoustic features}

In our experiments, we randomly choose training and evaluation data from $Fisher$ dataset, following \cite{lilantian2017deep}.
The training dataset consists of 5000 speakers randomly chosen from the $Fisher$ dataset, including 2500 male and 2500 female. This training dataset is used to train i-vector extractor, Kaldi xvector network, PLDA and our own network.
The evaluation dataset consists of 500 female and 500 male speakers randomly chosen from $Fisher$ dataset.
	There is no overlap in speakers between training and evaluation data.



The acoustic features are 23 dimensional MFCCs with a frame-length of 25ms. Mean-normalization over a sliding window of up to 3 seconds is performed on each dimension of the MFCC features. And an energy-based VAD is used to detect speech frames. All experiments are conducted on the detected speech frames.

\subsection{Baseline systems}
Two baseline systems are evaluated.
The first baseline is a traditional GMM-UBM i-vector system, which is based on the Kaldi recipe. 
Delta and acceleration are appended to obtain 69 dimensional feature vectors. The UBM is a 2048 component full-covariance GMM. The i-vector dimension is set to be 600. 

The second baseline is the Kaldi xvector system, which is detailed in the paper \cite{snyder2017deep} and the Kaldi toolkit. We use the default setting in the Kaldi script. 
Basically, the system use a feed-forward deep neural network with a temporal pooling layer that aggregates over the input speech frames. This enables the network to be trained (based on the softmax loss) to extract a fix-dimensional speaker embedding vector (called xvector) from a varying-duration speech segment.

\subsection{PLDA back-end}
After extracting the speaker embedding vectors, we need a scoring module, or say a back-end, to make verification decision.
Cosine distance and Euclidean distance are classic back-ends. Recently, likelihood-ratio score based back-ends such as PLDA (probabilistic linear discriminant analysis) have been shown to achieve superior performance.
In \cite{wang2017joint}, PLDA and various related models are compared.
For consistent comparisons, Kaldi's implementation of PLDA, including length normalization but without LDA dimensionality reduction, is used in all PLDA-related experiments in this paper.

\begin{table}[h]
	\centering
	\caption{Details of our network architecture. Numbers in parentheses denote the input and output dimensions in each layer. TDNN is time-delayed neural network. FC is fully connected neural network.}
	\label{tab:one}
	\begin{tabular}{|l|l|}
		\hline
		\multirow{2}{*}{utterance level layers} & FC\_2 (512$\to$300)      \\ \cline{2-2} 
		& FC\_1 (3000$\to$512)     \\ \hline\hline
		statistic pooling layer                 & mean and standard deviation     \\ \hline\hline
		\multirow{5}{*}{frame level layers}     & TDNN\_5 (512$\times$1$\to$1500) \\ \cline{2-2} 
		& TDNN\_4 (512$\times$1$\to$512)  \\ \cline{2-2} 
		& TDNN\_3 (512$\times$3$\to$512)  \\ \cline{2-2} 
		& TDNN\_2 (512$\times$3$\to$512)  \\ \cline{2-2} 
		& TDNN\_1 (23$\times$5$\to$512)   \\ \hline
	\end{tabular}
\end{table}

\subsection{Our network architecture}
Basically, we employ the same network architecture to generate speaker embedding vectors as in Kaldi's xvector recipe.
There are two minor differences in experiments.
First, we do not use natural gradient descent \cite{povey2014parallel} to optimize the network. Instead, we use the classic stochastic gradient descent.
In our experiments, the minibatch size is 1000 chunks, and each chunk contains 200 speech frames. The learning rate is initialized as 0.01 and then is multiplied by 0.9 after each epoch. The training stops when the learning rate drops to be below 0.0001, which roughly corresponds to 100 epochs of training.
Second, we use ReLU layer \cite{glorot2011deep} after batch normalization layer \cite{ioffe2015batch}, which is found to be more stable in training than using the two layers in the opposite order as employed in the Kaldi xvector network. 
Details of our network architecture are shown in Table \ref{tab:one}.

\section{Experimental Results}

\subsection{Experiment with fixed-duration enroll utterances}

\begin{figure}[h]
	\centering
	\includegraphics[width=8.6cm]{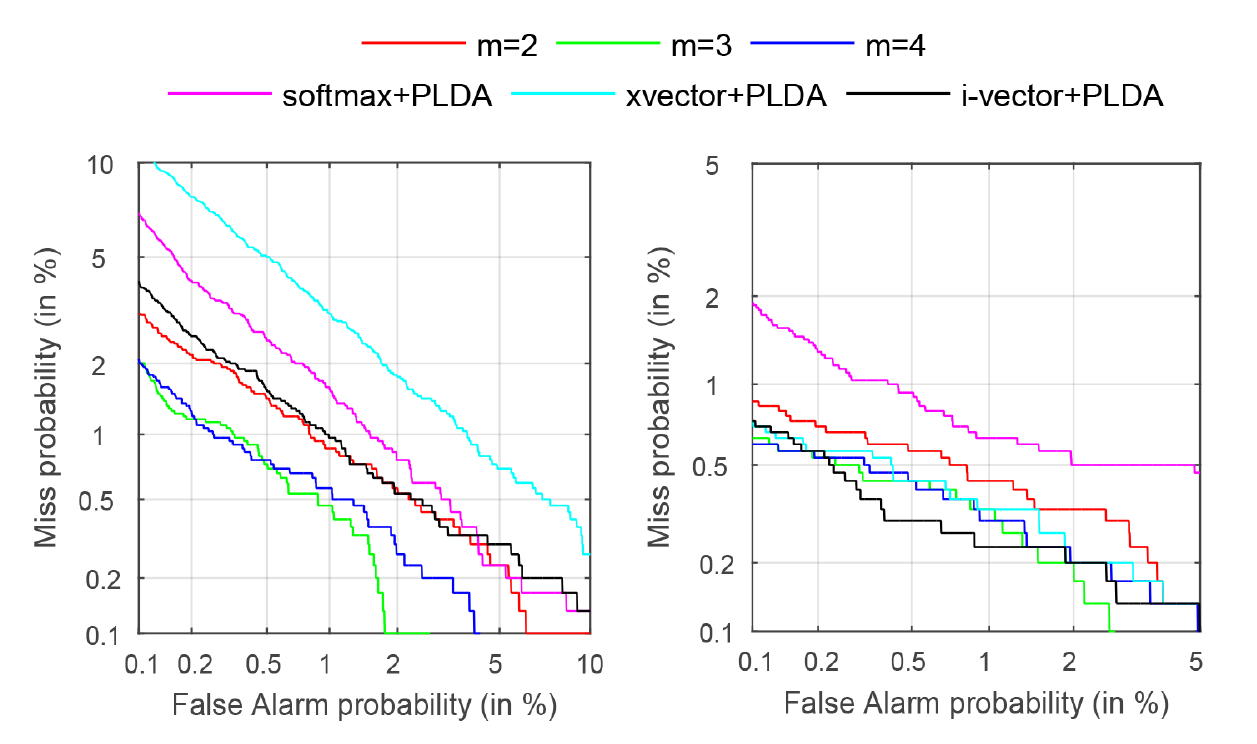}
	\caption{DET curves 
		with 3000-frame enrollment, under 300-frame test condition (left) and 1500-frame test condition (right).
The models can be our network ($m=2,3,4$ and ``softmax+PLDA''), i-vector and xvector.
}
	\label{fig:2}
	\vskip -0.2in
\end{figure}

In the first experiment, we fix the durations of enroll utterances to be 3000 frames after VAD. The durations of test utterances vary in $\{300,500,1000,1500\}$ frames after VAD. We choose 1 enroll utterance and 3 test utterances per speaker from the evaluation dataset. Together we have $1,000\times3,000=3,000,000$ trials, which consist of $3,000$ target trials and $2,997,000$ non-target trials. The results are given in Table \ref{tab2} and Figure \ref{fig:2}, which shows the effect of different test durations on speaker verification performance in the long enrollment case.

Some main observations are as follows.
\emph{First}, using triplet loss yields inferior performance. We follow the triplet sampling strategy in \cite{bredin2017tristounet}, which is also time consuming. 

\emph{Second}, for short test condition (300 and 500 frames), A-softmax performs significantly better than both i-vector and xvector baseline. 
When testing with longer utterances (1500 frames), i-vector system performs better, which is also observed in \cite{snyder2016deep,lilantian2017deep}.
Similar observations can be seen from Figure \ref{fig:2}, which shows the DET curves under 300 and 1500-frame test conditions.

\emph{Third}, to preclude the effect of the differences in network architecture in xvector system and our network, we can compare the results from softmax and A-softmax, both using our own network.
A-softmax outperforms traditional softmax significantly.
Compared to softmax with PLDA back-end, A-softmax with $m=3$ and cosine back-end achieves $48.46\%$, $58.76\%$, $47.14\%$ and $41.10\%$ EER relative reductions under 300, 500, 1000 and 1500-frame test conditions, respectively.

\emph{Forth}, ideally, larger angular margin $m$ could
bring stronger discrimination power. In practice, this does not always hold due to the complication of neural network training, as seen from Table \ref{tab2} and Figure \ref{fig:2}.


\begin{table}[h]
	\centering
	\caption{EERs (\%) for 3000-frame enroll utterances and different durations of test utterances. 
		m=2, 3, 4 are for A-softmax with m=2, 3, 4. 
		Cosine is for cosine distance. 
		Euclidean is for Euclidean distance. }
	\label{tab2}
	\begin{tabular}{l|l|cccc}
		\hline
		\multicolumn{2}{l|}{}                                                                                                                       & \multicolumn{4}{c}{Durations of test utterances} \\ \hline
		Model                                                                  & \begin{tabular}[c]{@{}l@{}}loss\\ (metric)\end{tabular}            & 300    & 500    & 1000   & 1500  \\ \hline\hline
		ivector                                                                & \begin{tabular}[c]{@{}l@{}}-\\ (PLDA)\end{tabular}                 & 1.00   & 0.53   & \textbf{0.33}   & \textbf{0.37}  \\ \hline
		xvector                                                                & \begin{tabular}[c]{@{}l@{}}softmax\\ (PLDA)\end{tabular}           & 1.86   & 0.83   & 0.40   & 0.43  \\ \hline\hline
		\multirow{3}{*}{\begin{tabular}[c]{@{}l@{}}our\\ network\end{tabular}} & \begin{tabular}[c]{@{}l@{}}softmax\\ (cosine)\end{tabular}         & 1.67   & 1.17   & 0.90   & 0.83  \\ \cline{2-6} 
		& \begin{tabular}[c]{@{}l@{}}softmax\\ (PLDA)\end{tabular}           & 1.30   & 0.97   & 0.70   & 0.73  \\ \cline{2-6} 
		& \begin{tabular}[c]{@{}l@{}}triplet loss\\ (Euclidean)\end{tabular} & 2.17   & 1.63   & 1.17   & 1.23  \\ \hline\hline
		\multirow{3}{*}{\begin{tabular}[c]{@{}l@{}}our\\ network\end{tabular}} & \begin{tabular}[c]{@{}l@{}}$m=2$\\ (cosine)\end{tabular}             & 0.94   & 0.60   & 0.47   & 0.57  \\ \cline{2-6} 
		& \begin{tabular}[c]{@{}l@{}}$m=3$\\ (cosine)\end{tabular}             & \textbf{0.67}   & \textbf{0.40}   & 0.37   & 0.43  \\ \cline{2-6} 
		& \begin{tabular}[c]{@{}l@{}}$m=4$\\ (cosine)\end{tabular}             & 0.70   & 0.47   & \textbf{0.33}   & 0.47  \\ \hline
	\end{tabular}
\end{table}

\begin{table}[h]
	\centering
	\caption{EERs (\%) with equal durations of enroll and test utterances.
}
	\label{tab3}
	\begin{tabular}{l|l|cccc}
		\hline
		\multicolumn{2}{l|}{}                                                                                                             & \multicolumn{4}{c}{Durations of utterances}                                                                          \\ \hline
		Model                                                                  & \begin{tabular}[c]{@{}l@{}}loss\\ (metric)\end{tabular}  & 300                      & 500                      & 1000                     & 1500                     \\ \hline\hline
		ivector                                                                & \begin{tabular}[c]{@{}l@{}}-\\ (PLDA)\end{tabular}       & 2.93                     & 1.57                     & \textbf{0.50}                     & \textbf{0.47}                     \\ \hline
		xvector                                                                & \begin{tabular}[c]{@{}l@{}}softmax\\ (PLDA)\end{tabular} & 3.17                     & 1.63                     & 0.63                     & 0.63                     \\ \hline
		\begin{tabular}[c]{@{}l@{}}our\\ network\end{tabular}                  & \begin{tabular}[c]{@{}l@{}}softmax\\ (PLDA)\end{tabular} & \multicolumn{1}{l}{3.43} & \multicolumn{1}{l}{2.40} & \multicolumn{1}{l}{1.20} & \multicolumn{1}{l}{1.07} \\ \hline\hline
		\multirow{6}{*}{\begin{tabular}[c]{@{}l@{}}our\\ network\end{tabular}} & \begin{tabular}[c]{@{}l@{}}$m=2$\\ (cosine)\end{tabular}   & 2.90                     & 1.57                     & 0.77                     & 0.83                     \\ \cline{2-6} 
		& \begin{tabular}[c]{@{}l@{}}$m=2$\\ (PLDA)\end{tabular}     & 2.17                     & 1.33                     & 0.73                     & 0.80                     \\ \cline{2-6} 
		& \begin{tabular}[c]{@{}l@{}}$m=3$\\ (cosine)\end{tabular}   & 2.50                     & \textbf{1.23}                     & 0.73                     & 0.56                     \\ \cline{2-6} 
		& \begin{tabular}[c]{@{}l@{}}$m=3$\\ (PLDA)\end{tabular}     & \textbf{2.10}                     & 1.33                     & 0.70                     & 0.77                     \\ \cline{2-6} 
		& \begin{tabular}[c]{@{}l@{}}$m=4$\\ (cosine)\end{tabular}   & 2.43                     & 1.33                     & 0.70                     & 0.63                     \\ \cline{2-6} 
		& \begin{tabular}[c]{@{}l@{}}$m=4$\\ (PLDA)\end{tabular}     & 2.23                     & 1.37                     & 0.73                     & 0.90                     \\ \hline
	\end{tabular}
\end{table}

\begin{figure}[h]
	\centering

	\includegraphics[width=8.6cm]{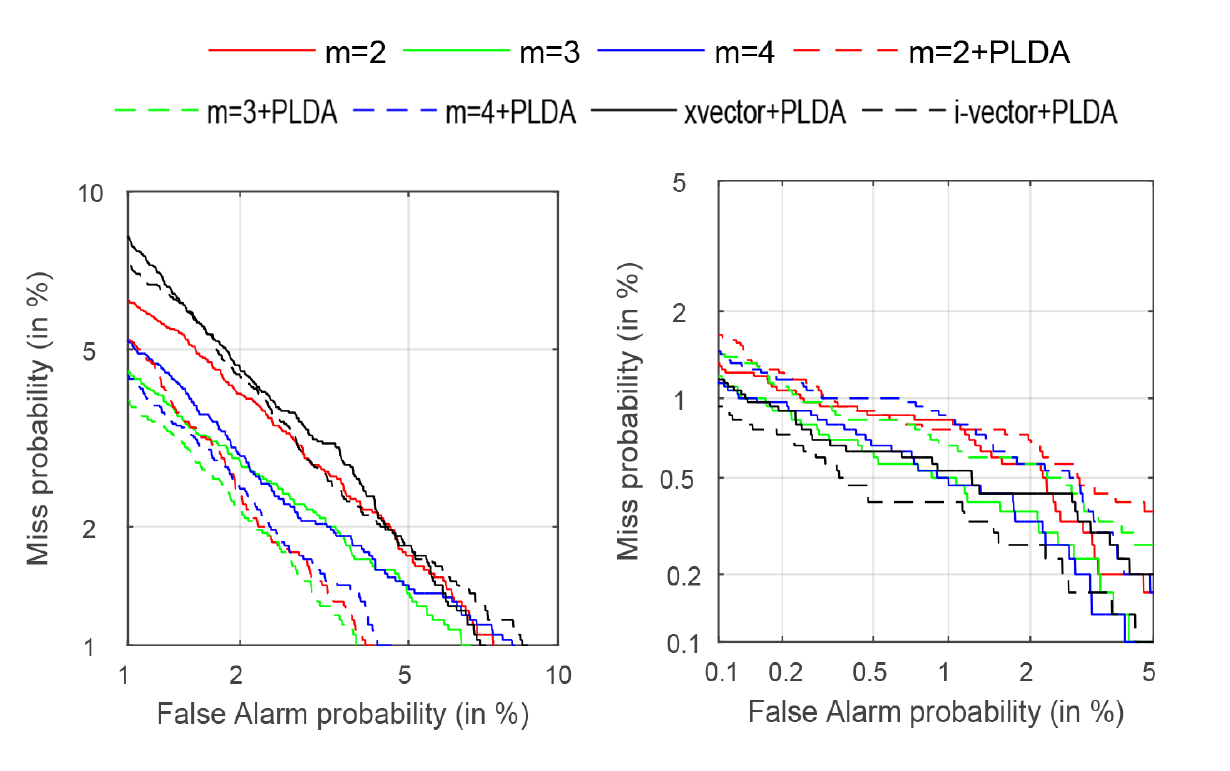}
	\caption{DET curves 
		with equal durations of enroll and test, 300-frame condition (left) and 1500-frame condition (right).
	}
	\label{fig:3}
\end{figure}

\subsection{Experiment with equal durations of enroll and test utterances}

In the second experiment, we set the durations of enroll and test utterances to be equal, varying in $\{300,500,1000,1500\}$ frames after VAD. 
The trials are created, following the same strategy as in the first experiment.
The results are given in Table \ref{tab3} and Figure \ref{fig:3}, which show the effect of different enroll and test durations on speaker verification performance.

Some main observations are as follows.
\emph{First}, in this experiment, we add the results of A-softmax with PLDA back-end, which should be compared to A-softmax with cosine back-end.
For short utterance condition (short in both enrollment and test, with 300 frames), using PLDA back-end significantly reduce EERs of the A-softmax systems.
For $m=2,3,4$, the EER relative reductions are $25.17\%$, 
$16.00\%$ and $8.23\%$ respectively.
For long utterance condition (long in both enrollment and test), using PLDA back-end does not always improve the A-softmax systems.
A possible reason is that during training, we slice the utterances into 200-frame chunks. 
Both the network and the PLDA are trained over 200-frame chunks, which consequently work best for short utterances.

\emph{Second}, we do not include the inferior result of triplet loss in Table \ref{tab3}.
Compared to the i-vector and xvector baseline, the EERs of A-softmax system are the best on almost all conditions, except that both the enroll and the test utterances are long (1000 and 1500 frames). This agree with the result in the first experiment and also in other papers \cite{snyder2016deep,lilantian2017deep}.

\emph{Third}, when comparing softmax and A-softmax, both using our own network, A-softmax outperforms traditional softmax significantly on all conditions.

%

\section{Conclusions}

In this work, we introduce A-softmax loss into end-to-end speaker verification, which outperforms softmax and triplet loss significantly, under the same neural network architecture.
Furthermore, we use PLDA as back-end to improve A-softmax under short utterance condition. 
Compared with Kaldi i-vector baseline, A-softmax achieves better performance except for long utterance condition.



\bibliographystyle{IEEEtran}

\bibliography{mybib}

\begin{thebibliography}{10}
\providecommand{\url}[1]{#1}
\csname url@samestyle\endcsname
\providecommand{\newblock}{\relax}
\providecommand{\bibinfo}[2]{#2}
\providecommand{\BIBentrySTDinterwordspacing}{\spaceskip=0pt\relax}
\providecommand{\BIBentryALTinterwordstretchfactor}{4}
\providecommand{\BIBentryALTinterwordspacing}{\spaceskip=\fontdimen2\font plus
\BIBentryALTinterwordstretchfactor\fontdimen3\font minus
  \fontdimen4\font\relax}
\providecommand{\BIBforeignlanguage}[2]{{%
\expandafter\ifx\csname l@#1\endcsname\relax
\typeout{** WARNING: IEEEtran.bst: No hyphenation pattern has been}%
\typeout{** loaded for the language `#1'. Using the pattern for}%
\typeout{** the default language instead.}%
\else
\language=\csname l@#1\endcsname
\fi
#2}}
\providecommand{\BIBdecl}{\relax}
\BIBdecl

\bibitem{dehak2011front}
N.~Dehak, P.~J. Kenny, R.~Dehak, P.~Dumouchel, and P.~Ouellet, ``Front-end
  factor analysis for speaker verification,'' \emph{IEEE Transactions on Audio,
  Speech, and Language Processing}, vol.~19, no.~4, pp. 788--798, 2011.

\bibitem{prince2007probabilistic}
S.~J. Prince and J.~H. Elder, ``Probabilistic linear discriminant analysis for
  inferences about identity,'' in \emph{ICCV}, 2007.

\bibitem{zhang2016end}
S.-X. Zhang, Z.~Chen, Y.~Zhao, J.~Li, and Y.~Gong, ``End-to-end attention based
  text-dependent speaker verification,'' in \emph{IEEE Spoken Language
  Technology Workshop (SLT)}, 2016.

\bibitem{heigold2016end}
G.~Heigold, I.~Moreno, S.~Bengio, and N.~Shazeer, ``End-to-end text-dependent
  speaker verification,'' in \emph{ICASSP}, 2016.

\bibitem{chowdhury2017attention}
F.~Chowdhury, Q.~Wang, I.~L. Moreno, and L.~Wan, ``Attention-based models for
  text-dependent speaker verification,'' \emph{arXiv preprint
  arXiv:1710.10470}, 2017.

\bibitem{li2017deep}
C.~Li, X.~Ma, B.~Jiang, X.~Li, X.~Zhang, X.~Liu, Y.~Cao, A.~Kannan, and Z.~Zhu,
  ``Deep speaker: an end-to-end neural speaker embedding system,'' \emph{arXiv
  preprint arXiv:1705.02304}, 2017.

\bibitem{bredin2017tristounet}
H.~Bredin, ``Tristounet: triplet loss for speaker turn embedding,'' in
  \emph{ICASSP}, 2017.

\bibitem{lilantian2017deep}
L.~Li, Y.~Chen, Y.~Shi, Z.~Tang, and D.~Wang, ``Deep speaker feature learning
  for text-independent speaker verification,'' in \emph{Interspeech}, 2017.

\bibitem{torfi2017text}
A.~Torfi, N.~M. Nasrabadi, and J.~Dawson, ``Text-independent speaker
  verification using 3d convolutional neural networks,'' in \emph{IEEE
  International Conference on Multimedia and Expo (ICME)}, 2017.

\bibitem{zhang2017end}
C.~Zhang and K.~Koishida, ``End-to-end text-independent speaker verification
  with triplet loss on short utterances,'' in \emph{Interspeech}, 2017.

\bibitem{wan2017generalized}
L.~Wan, Q.~Wang, A.~Papir, and I.~L. Moreno, ``Generalized end-to-end loss for
  speaker verification,'' \emph{arXiv preprint arXiv:1710.10467}, 2017.

\bibitem{snyder2016deep}
D.~Snyder, P.~Ghahremani, D.~Povey, D.~Garcia-Romero, Y.~Carmiel, and
  S.~Khudanpur, ``Deep neural network-based speaker embeddings for end-to-end
  speaker verification,'' in \emph{IEEE Spoken Language Technology Workshop
  (SLT)}, 2016.

\bibitem{schroff2015facenet}
F.~Schroff, D.~Kalenichenko, and J.~Philbin, ``Facenet: A unified embedding for
  face recognition and clustering,'' in \emph{IEEE Conference on Computer
  Vision and Pattern Recognition (CVPR)}, 2015.

\bibitem{liu2017sphereface}
W.~Liu, Y.~Wen, Z.~Yu, M.~Li, B.~Raj, and L.~Song, ``Sphereface: Deep
  hypersphere embedding for face recognition,'' in \emph{IEEE Conference on
  Computer Vision and Pattern Recognition (CVPR)}, 2017.

\bibitem{snyder2017deep}
D.~Snyder, D.~Garcia-Romero, D.~Povey, and S.~Khudanpur, ``Deep neural network
  embeddings for text-independent speaker verification,'' in
  \emph{Interspeech}, 2017.

\bibitem{Povey_ASRU2011}
D.~Povey, A.~Ghoshal, G.~Boulianne, L.~Burget, O.~Glembek, N.~Goel,
  M.~Hannemann, P.~Motlicek, Y.~Qian, P.~Schwarz, J.~Silovsky, G.~Stemmer, and
  K.~Vesely, ``The kaldi speech recognition toolkit,'' in \emph{IEEE Workshop
  on Automatic Speech Recognition and Understanding (ASRU)}, 2011.

\bibitem{wang2017joint}
Y.~Wang, H.~Xu, and Z.~Ou, ``Joint bayesian gaussian discriminant analysis for
  speaker verification,'' in \emph{ICASSP}, 2017.

\bibitem{povey2014parallel}
D.~Povey, X.~Zhang, and S.~Khudanpur, ``Parallel training of deep neural
  networks with natural gradient and parameter averaging,'' in \emph{ICLR},
  2014.

\bibitem{glorot2011deep}
X.~Glorot, A.~Bordes, and Y.~Bengio, ``Deep sparse rectifier neural networks,''
  in \emph{AISTATS}, 2011.

\bibitem{ioffe2015batch}
S.~Ioffe and C.~Szegedy, ``Batch normalization: Accelerating deep network
  training by reducing internal covariate shift,'' in \emph{ICML}, 2015.

\end{thebibliography}


\end{document}